%% file: main_body.tex
\pdfoutput=1
\PassOptionsToPackage{unicode}{hyperref}
\PassOptionsToPackage{hyphens}{url}
\PassOptionsToPackage{dvipsnames,svgnames,x11names}{xcolor}
\documentclass[12pt]{article}

\usepackage{amsmath,amssymb,amsthm,bm}

\usepackage{setspace}
\usepackage{threeparttable}
\usepackage{rotating}
\usepackage{multirow}
\usepackage{makecell}
\usepackage[lined,boxed,commentsnumbered,ruled,vlined]{algorithm2e}
\usepackage{xr-hyper}

\usepackage{iftex}
\usepackage[T1]{fontenc}
\usepackage[utf8]{inputenc}
\usepackage{textcomp}
\usepackage{lmodern}
\IfFileExists{upquote.sty}{\usepackage{upquote}}{}
\IfFileExists{microtype.sty}{%
  \usepackage[]{microtype}
  \UseMicrotypeSet[protrusion]{basicmath}
}{}
\makeatletter
\@ifundefined{KOMAClassName}{%
  \IfFileExists{parskip.sty}{\usepackage{parskip}}{%
    \setlength{\parindent}{0pt}
    \setlength{\parskip}{6pt plus 2pt minus 1pt}}
}{\KOMAoptions{parskip=half}}
\makeatother
\usepackage{xcolor}
\makeatletter
\ifx\paragraph\undefined\else
  \let\oldparagraph\paragraph
  \renewcommand{\paragraph}{%
    \@ifstar\xxxParagraphStar\xxxParagraphNoStar}
  \newcommand{\xxxParagraphStar}[1]{\oldparagraph*{#1}\mbox{}}
  \newcommand{\xxxParagraphNoStar}[1]{\oldparagraph{#1}\mbox{}}
\fi
\ifx\subparagraph\undefined\else
  \let\oldsubparagraph\subparagraph
  \renewcommand{\subparagraph}{%
    \@ifstar\xxxSubParagraphStar\xxxSubParagraphNoStar}
  \newcommand{\xxxSubParagraphStar}[1]{\oldsubparagraph*{#1}\mbox{}}
  \newcommand{\xxxSubParagraphNoStar}[1]{\oldsubparagraph{#1}\mbox{}}
\fi
\makeatother

\usepackage{longtable,booktabs,array}
\usepackage{calc}
\usepackage{etoolbox}
\makeatletter
\patchcmd\longtable{\par}{\if@noskipsec\mbox{}\fi\par}{}{}
\makeatother
\IfFileExists{footnotehyper.sty}{\usepackage{footnotehyper}}{\usepackage{footnote}}
\makesavenoteenv{longtable}
\usepackage{graphicx}
\makeatletter
\def\maxwidth{\ifdim\Gin@nat@width>\linewidth\linewidth\else\Gin@nat@width\fi}
\def\maxheight{\ifdim\Gin@nat@height>\textheight\textheight\else\Gin@nat@height\fi}
\makeatother
\setkeys{Gin}{width=\maxwidth,height=\textheight,keepaspectratio}
\graphicspath{{Fig/}}
\makeatletter
\def\fps@figure{htbp}
\makeatother

\makeatletter
\@ifpackageloaded{caption}{}{\usepackage{caption}}
\AtBeginDocument{%
\ifdefined\contentsname\renewcommand*\contentsname{Table of contents}\else\newcommand\contentsname{Table of contents}\fi
\ifdefined\listfigurename\renewcommand*\listfigurename{List of Figures}\else\newcommand\listfigurename{List of Figures}\fi
\ifdefined\listtablename\renewcommand*\listtablename{List of Tables}\else\newcommand\listtablename{List of Tables}\fi
\ifdefined\figurename\renewcommand*\figurename{Figure}\else\newcommand\figurename{Figure}\fi
\ifdefined\tablename\renewcommand*\tablename{Table}\else\newcommand\tablename{Table}\fi}
\@ifpackageloaded{float}{}{\usepackage{float}}
\floatstyle{ruled}
\@ifundefined{c@chapter}{\newfloat{codelisting}{h}{lop}}{\newfloat{codelisting}{h}{lop}[chapter]}
\floatname{codelisting}{Listing}

\makeatother
\makeatletter
\@ifpackageloaded{caption}{}{\usepackage{caption}}
\@ifpackageloaded{subcaption}{}{\usepackage{subcaption}}
\makeatother

\ifLuaTeX
  \usepackage{selnolig}
\fi
\usepackage[authoryear,round]{natbib}
\usepackage{hyperref}
\usepackage{bookmark}
\IfFileExists{xurl.sty}{\usepackage{xurl}}{}
\hypersetup{
  pdftitle={Bayesian Simultaneous Credible Bands for Polynomial Regression},
  pdfauthor={Fei Yang; Yang Han; Wei Liu},
  pdfkeywords={Bayesian simultaneous credible bands, dose response study, multiple testing, simultaneous inference, polynomial regression},
  colorlinks=true,
  linkcolor={blue},
  filecolor={Maroon},
  citecolor={Blue},
  urlcolor={Blue},
  pdfcreator={LaTeX via pandoc}}

\newcommand{\anon}{1}

\theoremstyle{plain}
\newtheorem{theorem}{Theorem}[section]

\theoremstyle{definition}

\begin{document}

\def\spacingset#1{\renewcommand{\baselinestretch}{#1}\small\normalsize}
\spacingset{1}

\if1\anon
{
  \title{\bf Bayesian Simultaneous Credible Bands for Polynomial Regression}
  \author{Fei Yang$^{1}$, Yang Han$^{1}$\thanks{Corresponding author. E-mail: \href{mailto:yang.han@manchester.ac.uk}{yang.han@manchester.ac.uk}.}, Wei Liu$^{2}$, Ian Hall$^{1}$   \\[6pt]
  $^{1}$Department of Mathematics, The University of Manchester, UK \\
  $^{2}$School of Mathematical Sciences and \\Southampton Statistical Sciences Research Institute,\\ University of Southampton, UK
}
 \maketitle
}
\fi

\if0\anon
{
  \bigskip
  \bigskip
  \bigskip
  \begin{center}
    {\LARGE\bf Bayesian Simultaneous Credible Bands for Polynomial Regression}
  \end{center}
  \medskip
}
\fi

\bigskip
\begin{abstract}

Quantifying efficacy uncertainty across the entire dose range is crucial in dose–response studies. Although the frequentist simultaneous confidence band (FSCB) is widely used for this purpose, it does not readily incorporate prior knowledge. The Bayesian simultaneous credible band (BSCB) offers a natural alternative, yet practical methods for constructing BSCBs remain scarce in the literature. In this paper, we propose a unified framework for constructing a BSCB for the regression curve in a univariate polynomial model over a finite covariate interval. An efficient simulation-based procedure is developed to determine the critical constant of a BSCB. The framework accommodates inference under different levels of prior information and can be implemented either analytically or via posterior sampling methods. Notably, we prove that under mild regularity conditions, the BSCB is asymptotically equivalent to the FSCB, thereby attaining the nominal frequentist coverage for a broad class of priors. Simulation studies confirm that the BSCB attains the exact posterior simultaneous coverage probability across various scenarios. An application to a dose–response study illustrates its importance in identifying the minimum effective dose in Phase II clinical trials. Software implementation of the proposed methods is available in an accompanying R package.

\end{abstract}

\vfill

\newpage
\spacingset{1.8} 

\section{Introduction}
\label{sect-intro}

Simultaneous confidence band (SCB) has emerged as a powerful tool for quantifying uncertainty in an unknown regression function over a given covariate range and thus plays a vital role in statistical inference.
Unlike the pointwise confidence band (PCB), which is formed by connecting the confidence interval at each covariate point, a ($1-\alpha$) SCB covers the entire regression function over a given covariate range $(a,b)$ with a confidence level of at least $1-\alpha$.
This property makes the SCB particularly valuable in the dose-response study, which is an integral component of clinical drug development. 
The goal is to quantify the uncertainty of the dose-response curve and assess efficacy for all patients across different dose levels with a guaranteed confidence level $1-\alpha$. 
This involves rigorous procedures in accordance with the International Council for Harmonisation (ICH) guidelines \citep{ICH1994}, as an insufficient dose may lead to inadequate efficacy, whereas an excessive dose can result in safety or tolerability issues. 
There are many choices to capture the dose-response relationship, among which the polynomial model is a commonly used and flexible one \citep*{bretz2005combining}.

Over the past several decades, frequentist simultaneous confidence bands (FSCBs) for the polynomial regression model have been extensively developed. Existing approaches can be broadly classified into three categories: 
(i) Numerical-integration-based methods \citep{Wynn1971, Spurrier01081993} provide exact or conservative bands but typically require certain assumptions on the design matrix; 
(ii) Tube-volume approaches \citep{Naiman1986, sun1994simultaneous} yield conservative bands for a general curvilinear regression model but require information on upper bounds for the roughness of the model; 
(iii) Simulation-based methods, which include bootstrap and numerical integration methods \citep*{kosorok1999exact,liu2008statistical, Liu2014}, can deliver exact bands but are often model-specific.

Nevertheless, these FSCBs face two main limitations. 
First, they do not incorporate prior information.
In clinical trials, leveraging expert knowledge from historical data can substantially reduce the sample size required to achieve the desired power, thereby lowering the overall cost \citep{schmidli2014robust}.
The Bayesian framework is particularly well-suited to this objective, allowing inference to be updated as new information accrues across different clinical stages.
Second, the frequentist interpretation of coverage as a long-term frequency over repeated experiments under identical conditions may be difficult to satisfy due to unforeseen factors in practice.
In contrast, the Bayesian approach is more intuitive in providing a direct probability statement about uncertainty given the observed data from a single experiment and prior information.

The Bayesian simultaneous credible band (BSCB) offers an alternative that addresses these limitations, yet it has received comparatively less attention.
\citet{liu2010} very briefly mentioned a Bayesian approach for linear regression but provided no details. 
Furthermore, \citet{besag1995bayesian} and
\citet{montiel2019simultaneous} constructed BSCBs as products of univariate intervals, which are restricted to hyperrectangular regions and only ensure simultaneous coverage over a finite set of covariate points rather than a continuous covariate interval $(a,b)$. 
\citet{held2004simultaneous} derived simultaneous posterior probability statements via Rao-Blackwellization, but with a substantial computational cost for large samples.
Related methods for nonparametric models (\citealp{crainiceanu2007spatially}; \citealp*{krivobokova2010simultaneous}) and latent Gaussian models \citep{sorbye2011simultaneous} are also available, but they are conservative and do not attain exact nominal coverage $1-\alpha$.

To address the aforementioned gaps, we propose a novel framework for constructing a $(1-\alpha)$ level two-sided exact BSCB for the polynomial model over the given covariate interval $(a,b)$. Three approaches have been developed based on (i) the normal-gamma conjugate prior, (ii) the independent Jeffreys prior for objective inference, and (iii) the non-conjugate prior implemented by the
Hamiltonian Monte Carlo (HMC) method. 
To the best of our knowledge, the proposed methods appear to be the first to achieve exact nominal posterior simultaneous coverage $1-\alpha$ for polynomial models. 

%Contribution
This paper makes three main contributions. 
First, we develop a new methodology to construct the exact BSCB, which is flexible and applies to both the i.i.d. error case and the autoregressive error structure where the autoregressive coefficient is assumed to be known.
The framework accommodates inference under different prior specifications tailored to the observed dataset.
Second, we introduce an efficient computational algorithm for determining the critical constant of a BSCB.
Third, we show that, under the independent Jeffreys prior, the BSCB is identical to the FSCB in finite samples.
Furthermore, the general BSCBs have asymptotically frequentist properties for attaining empirical coverage as the sample size $n \to \infty$ under mild regularity conditions.

%Paper Structure
Accordingly, this paper is organized as follows. 
Section \ref{sect-methods} introduces the proposed BSCBs under three types of priors, together with the computation of the critical constants and a concrete comparison with the frequentist simultaneous bands, the frequentist pointwise bands, and the Bayesian pointwise bands.
In Section \ref{sect-theory}, we establish the finite-sample equivalence between BSCB and FSCB under the independent Jeffreys prior and the asymptotic equivalence more generally.
Section \ref{sect-simulation} evaluates the performance of BSCBs through simulation studies for the polynomial regression models with independent random error and autoregressive error.
Section \ref{sect-realexample} illustrates the practical feasibility of the proposed methods to determine the minimum effective dose using a phase II study from \citet{bretz2005combining}.
Final conclusions are presented in Section \ref{sect-conclusion}.

\section{Methodology}\label{sect-methods}

\subsection{Preliminaries}\label{subsect-preliminaries}

Consider a $p$-th order polynomial regression model $Y=\theta_0+\theta_1x+\dots +\theta_p x^p +e$, $p\ge 2$, with observations 
\begin{equation}
\bm{Y}=\bm{X}\bm{\theta}+\bm{e},
\label{eq_5.2.1}
\end{equation}
where $\bm{Y}=(y_1,\dots,y_n)^T$ is the $n$ dimensional vector of observed responses
corresponding to the covariate values $(x_1, \ldots, x_n)$, the design matrix $\bm{X}$ is a $n\times(p+1)$ full column-rank matrix with the  $l$-th $(1\le l \le n)$ row given by $(1, x_l,\dots, x_l^p)$, 
$\bm{\theta}=(\theta_0,\dots,\theta_p)^T\in \mathbb{R}^{p+1}$ is the  vector of unknown regression coefficients, and $\bm{e}=(e_1,\dots, e_n)^T$ is a vector of random errors with $\bm{e} \sim N(0,\sigma^2V)$ where 
the covariance matrix $V$ is assumed to be a known positive-definite matrix, and $\sigma^2$ is an unknown parameter. 
The unknown parameters $\bm{\theta}$ and $\sigma^2$ characterize the model and are estimated from the observed data. 
In dose-response studies, ${Y}$ can be an efficacy or a safety variable observed at a given dose level $x$, including the placebo.

For any $x\in (a,b)$, denote $\bm{x} = (1,x,\dots,x^p)^T$. A SCB for the regression function $\bm{x}^T\bm{\theta}$ over a given covariate interval $(a,b)$ provides useful information more directly on the plausible range of the regression function than the point or confidence set estimates of the unknown regression coefficients $\bm{\theta}$ and variance $\sigma^2$. 

In this study, we consider the construction of $(1-\alpha)$ level BSCBs for $\bm{x}^T\bm{\theta}$ of the form:
\begin{equation}
    P \left \{ \bm{x}^T\bm{\theta}\in \bm{x}^TE(\bm{\theta}|\bm{Y}) \pm \lambda\sqrt{\text{Var}(\bm{x}^T\bm{\theta})} \quad \forall x \in (a,b) \right \}=1-\alpha.
    \label{eq2}
\end{equation}
Here, $P\{\cdot\}$, $E(\cdot)$, and $\text{Var}(\cdot)$ are with respect to the posterior distribution of $\bm{\theta}$ given $\bm{Y}$. The critical constant $\lambda$ is chosen so that the posterior probability of $\bm{x}^T\bm{\theta}$ being contained in the given intervals simultaneously for all $x\in(a,b)$ is equal to $1-\alpha$. It is clear that $\lambda$ governs the width of the BSCB, and 
\[
\begin{aligned}
     &P\{ \bm{x}^T\bm{\theta}  \in \bm{x}^TE(\bm{\theta}|\bm{Y})\pm \lambda\sqrt{\text{Var}(\bm{x}^T\bm{\theta})}  , \quad \forall x \in (a,b)\} \\
     = & P \left \{-\lambda \le \frac{\bm{x}^T(\bm{\theta}-E(\bm{\theta}|\bm{Y}))}{\sqrt{\text{Var}(\bm{x}^T\bm{\theta})}} \le \lambda , \quad \forall x \in (a,b) \right \} \\
     = & P \left \{ S \le \lambda \right \} \, ,
\end{aligned}
\]
where
\[S \, = \, \sup_{x \in (a,b)}\frac{ \left |\bm{x}^T(\bm{\theta}-E(\bm{\theta}
    |\bm{Y})) \right |}{\sqrt{\text{Var}(\bm{x}^T\bm{\theta})}} \, .\]
In order to determine the critical constant $\lambda$, 
one needs to evaluate the posterior distribution of $S$.
This requires specifying a prior distribution for $\bm{\theta}$ and $\sigma^2$ and computing the posterior distributions of $\bm{\theta}$ and then $S$.  

%This band is typically referred to as the "sup-t" bands \citep{montiel2019simultaneous}, which is trivially the narrowest simultaneous confidence band compared to other frequentist methods such as the pointwise method, Bonferroni method \citep{b2004b}, and $\text{\v{S}id\'ak}$ projection method \citep{vsidak1967rectangular}. 

\subsection{Posterior estimation}\label{subsect-posterior}

To accommodate different inferential scenarios, we develop three types of BSCB based on the following priors: 
the normal-gamma conjugate prior (BSCB-C), the independent Jeffreys prior (BSCB-I-J), and non-conjugate priors implemented via the HMC method (BSCB-H).

\subsubsection{Normal-gamma conjugate prior}\label{sect-conjugate prior}
In Bayesian analysis, the choice of prior can significantly influence parameter estimates, especially variance components \citep{Gelman2006prior}. 
When initial beliefs on $\bm{\theta}$ and $\sigma^2$ are available, the normal-gamma conjugate prior is one of the most popular conjugate priors for the polynomial model.
Let $\tau = \sigma^{-2}$ be the precision of the random errors. A normal-gamma conjugate prior for $(\bm{\theta}, \tau)$ is given by 
\[
\xi(\bm{\theta},\tau) = \xi_1(\bm{\theta}|\tau) \cdot \xi_2(\tau),\quad \bm{\theta}\in \mathbb{R}^{p+1},\tau>0,
\]
where $\xi_1(\bm{\theta}|\tau) \sim N(\bm{\mu}, (\tau\mathcal{P})^{-1})$ and $\xi_2(\tau) \sim \text{Gamma}(\alpha_0, \beta_0)$.
Here, $\bm{\mu}$ is a known $(p+1)$ dimensional column vector, $\mathcal{P}$ is a known $(p+1)\times (p+1)$ positive definite matrix,  $\alpha_0>0$ is the known shape parameter, and  $\beta_0>0$ is the known rate parameter. The density function of $\text{Gamma}(\alpha_0, \beta_0)$ is $\xi_2(\tau)=\frac{\beta_0^{\alpha_0}}{\Gamma(\alpha_0)}\tau^{(\alpha_0-1)}\exp\{-\beta_0\tau\}$.
By combining the prior with the likelihood function, we obtain the joint posterior density $\xi(\bm{\theta},\tau|\bm{Y})$ which follows a normal-gamma distribution (and the reader is referred to, e.g., \citet{broemeling2017bayesian} pp. 1-3 for details):
\[
\begin{aligned}
&    \xi(\bm{\theta},\tau|\bm{Y})\sim \text{Normal-Gamma}(\bm{\theta},\tau|\bm{\mu}_n,\mathcal{P}_n,\alpha_n,\beta_n)
    \label{NG_prior} \\
& i.e.\ \  \bm{\theta}|\tau,\bm{Y} \sim N(\bm{\mu}_n, \tau^{-1}\mathcal{P}_n^{-1}) \ \ \ \text{and} \ 
    \tau|\bm{Y} \sim \text{Gamma}(\alpha_n, \beta_n),
\end{aligned}    
\]
where $\bm{\mu}_n=\mathcal{P}_n^{-1}(\bm{X}^TV^{-1}\bm{Y}+\mathcal{P}\bm{\mu})$, $\mathcal{P}_n=\bm{X}^TV^{-1}\bm{X}+\mathcal{P}$, $\alpha_n=(n+2\alpha_{0})/2$ and $\beta_n = \beta_{0}+(c/2)$. 
Here, $\alpha_n$ is the updated shape parameter, $\beta_n$ is the updated rate parameter, and $c=\bm{Y}^TV^{-1}\bm{Y}+\bm{\mu}^T\mathcal{P}\bm{\mu}-\bm{\mu}_n^T\mathcal{P}_n\bm{\mu}_n$ is a constant.

Our primary interest is the marginal posterior density of $\bm{\theta}$,  which is obtained by integrating out $\tau$ and given by the multivariate $t$ distribution \citep{genz2009computation}:
\begin{equation}
    \xi_1(\bm{\theta}|\bm{Y}) \sim t_{\nu_{\text{NG}}}(\bm{\mu}_{\text{NG}},\Sigma_{\text{NG}}),
    \label{NG_theta_posterior}
\end{equation}
where degrees of freedom $\nu_{\text{NG}}$, location vector $\bm{\mu}_{\text{NG}}$ and scale matrix  $\Sigma_{\text{NG}}$ in the form of 
\[
\begin{aligned}
    \nu_{\text{NG}}&=2\alpha_n=2\alpha_0+n,\\
    \bm{\mu}_{\text{NG}}&=\bm{\mu}_n =(\bm{X}^TV^{-1}\bm{X}+\mathcal{P})^{-1}(\bm{X}^TV^{-1}\bm{Y}+\mathcal{P}\bm{\mu}),\\
    \Sigma_{\text{NG}} &=  \mathcal{P}_n^{-1}(\beta_n/\alpha_n)=(\bm{X}^TV^{-1}\bm{X}+\mathcal{P})^{-1}\frac{2\beta_0+c}{2\alpha_0+n}.
\end{aligned}
\]

Hyperparameters for the conjugate prior may be elicited from expert knowledge or historical data \citep{kadane1980interactive}. 
In the absence of such information, several weakly informative specifications may be adopted, including the empirical Bayes approach, unit-information priors \citep{kass1995reference}, and g-priors \citep{zellner1986assessing}, see Section~\ref{subsecct-comp}.

To construct the BSCB in formula (\ref{eq2}), we require the posterior mean and variance of $\bm{\theta}$.
Based on the properties of the multivariate $t$ distribution in (\ref{NG_theta_posterior}), the posterior mean is $E(\bm{\theta}|\mathbf{Y})=\bm{\mu}_{\text{NG}}$ (for $\nu_{\text{NG}}>1$), which serves as the Bayes estimator and asymptotically approaches the maximum likelihood estimator (MLE).
Define $\nu_{\text{NG}}^{*}=\nu_{\text{NG}}/(\nu_{\text{NG}}-2)$.
The posterior covariance is $\text{Cov}(\bm{\theta}|\mathbf{Y}) = \nu_{\text{NG}}^{*}\Sigma_{\text{NG}}$ (for $\nu_{\text{NG}}>2$).
Consequently, the required variance for the regression function is $\text{Var}(\bm{x}^T\bm{\theta}|\mathbf{Y}) = \nu_{\text{NG}}^{*}\bm{x}^T\Sigma_{\text{NG}}\bm{x}$. 
Details on the derivation for the case of $V = I_n$ can be found in \citet{broemeling2017bayesian}, while the case of a known positive-definite matrix $V$ considered in this paper can be dealt with only minor modifications.

\subsubsection{Independent Jeffreys prior}\label{sect-non-informative prior}
In situations where little prior information is available, a non-informative prior is often employed. 
Here we consider the independent Jeffreys prior \citep{jeffreys1961theory}, which is invariant under monotone transformations of the parameter and regarded as the standard choice for location-scale problems \citep*{berger2024objective}.

Following \citet{gelman2013bayesian}, we assign a flat prior to $\bm{\theta}$ and the Jeffreys prior to $\tau$,  that is, $\xi(\bm{\theta}) \propto 1$ and $\xi(\tau)\propto [I(\tau)]^{1/2} \propto \tau^{-1}$, where $I(\tau)$ denotes the Fisher information.
Under the assumption of prior independence, the joint prior density is
\begin{equation*}
    \xi(\bm{\theta}, \tau) =\xi(\bm{\theta})\xi(\tau) \propto \tau^{-1}.
\end{equation*}
This non-informative prior avoids favoring any particular value of parameters and can be viewed as a limiting case of the conjugate prior in Section \ref{sect-conjugate prior} as the hyperparameters $\alpha_0,\beta_0 \to 0$ and $\mathcal{P}\to \bm{0}_{(p+1)\times (p+1)}$. 
Consequently, we can obtain the posterior distribution as in Section \ref{sect-conjugate prior}. 
Specifically, we have 
\[
\bm{\theta}|\tau,\bm{Y}\sim N(\hat{\bm{\theta}}_{\text{GLS}},\tau^{-1}\Sigma_0) \quad \text{and} \quad \tau|\bm{Y} \sim \text{Gamma}(\frac{\nu_J}{2}, \frac{\text{SSE}}{2}),
\]
where $\Sigma_0=(\bm{X}^TV^{-1}\bm{X})^{-1}$, $\text{SSE}=\nu_J\hat{\sigma}_{\text{GLS}}^2=  (\bm{Y-\bm{X}\hat{\bm{\theta}}_{\text{GLS}}})^{T}V^{-1}(\bm{Y-\bm{X}\hat{\bm{\theta}}_{\text{GLS}}})$, 
$\hat{\bm{\theta}}_{\text{GLS}}=(\bm{X}^TV^{-1}\bm{X})^{-1}\bm{X}^TV^{-1}\bm{Y}$,
and the degrees of freedom $\nu_J = n-p-1$.
Further, the posterior distribution of $\bm{\theta}$ given $\bm{Y}$ follows a multivariate $t$ distribution:
\begin{equation}
    \xi_1(\bm{\theta}|\bm{Y}) \sim t_{\nu_{J}}(\bm{\mu}_{J}, \Sigma_{J}),
  \label{eq_Jeffreys_posterior}
\end{equation}
with the location parameter of $\boldsymbol{\mu}_J = \hat{\bm{\theta}}_{\text{GLS}},$
and the scale matrix $\Sigma_J =\hat{\sigma}_{\text{GLS}}^2
(\mathbf{X}^T V^{-1}\mathbf{X})^{-1}.$
The posterior moments are subsequently given by $E(\bm{\theta}|\bm{Y}) = \bm{\mu}_{J}$ and $\text{Cov}(\bm{\theta}|\bm{Y}) = \Sigma_{J}\cdot \nu_J/(\nu_J-2)$.

An important property of the independent Jeffreys prior is that the resulting BSCB is the same as the FSCB in \citet{liu2008statistical} for finite sample sizes.
This alignment provides a frequentist validation for our Bayesian inference with finite samples, as further elaborated in Section~\ref{subsect-comparison} and Section~\ref{subsect_property_BSCB-I}.

\subsubsection{Non-conjugate priors implemented via HMC}\label{sub-sub-sect-HMC}
While the conjugate prior provides a closed-form posterior distribution, 
it relies on specific prior assumptions and may become restrictive when the prior moves beyond conjugacy. 
For example, the inverse-gamma prior on $\sigma^2$, equivalently the gamma prior on $\tau$, may be unsuitable when $\sigma$ is near zero \citep{Gelman2006prior}.
This motivates the use of more robust prior specifications.

When prior knowledge suggests a small $\sigma$, we consider two non-conjugate priors following \citet{Gelman2006prior}: a normal-half-normal prior and a normal-half-Cauchy prior. 
Specifically, independent Gaussian priors are assigned to the regression coefficients $\theta_j \sim N(0,(c_1\hat{\sigma}_{\theta_j})^2)$, where $\hat{\sigma}_{\theta_j}=\sqrt{\left[(\bm{X}^TV^{-1}\bm{X})^{-1}\right]_{jj}}$ is the 
square root of the $j$-th diagonal element in the covariance matrix.
For $\sigma$, we use either
a half-normal prior $N^{+}(0,(c_2\hat{\sigma}_{\text{GLS}})^2)$, or
a half-Cauchy prior $(0,s)$ with scale parameter $s$, where $c_1$ and $c_2$ control the prior dispersion.

Under these priors, the posterior distribution of $\bm{\theta}$ no longer has a well-known form, and HMC is therefore used to generate posterior samples.
We run four independent chains with 8000 iterations each, discard the first half as burn-in, and retain one sample out of every four,
which gives a total of $N=4000$ posterior samples. 
Following \citet{gelman2013bayesian}, the thinning lag is chosen to be larger than the lag at which the sample autocorrelation becomes negligible; see detailed discussions in Web Appendix~\ref{Appendix_Simulation_HMC}.
We also use the  No-U-Turn Sampler (NUTS) algorithm \citep{hoffman2014no}, with maximum tree depth of 15 and target acceptance rate of $0.95$ for better sampling performance based on our numerical investigation.
After obtaining the posterior samples  $\{\bm{\theta}^{(i)}\}_{i=1}^N$,  we obtain the estimated expectation $\hat{E}(\bm{\theta}|\bm{Y})=\frac{1}{N}\sum_{i=1}^N\bm{\theta}^{(i)}$
and covariance $\widehat{\text{Cov}}(\bm{\theta}|\bm{Y})=\frac{1}{N-1}\sum_{i=1}^N(\bm{\theta}^{(i)}-\hat{E}(\bm{\theta}|\bm{Y}))(\bm{\theta}^{(i)}-\hat{E}(\bm{\theta}|\bm{Y}))^T$. 

\subsection{Computation of the critical constant}\label{subsect-critical}

Since the expectation $E(\bm{\theta}|\bm{Y})$
and covariance $\text{Cov}(\bm{\theta}|\bm{Y})$ could be obtained either analytically or by HMC sampling, the only source of randomness in
expression (\ref{eq2}) is from $\bm{\theta}$, which follows its posterior distribution.
Recall that 
$
P\{S \le \lambda\} = 1-\alpha
$
where 
\begin{equation}
    S=\sup_{x \in (a,b)} T(x)
    \label{eq10}
\end{equation}
with $T(x)=|\bm{x}^T(\bm{\theta}-E(\bm{\theta}|\bm{Y}))|/\sqrt{\bm{x}^T\text{Cov}(\bm{\theta}|\bm{Y})\bm{x}}$. Hence the critical constant $\lambda$ is the $(1-\alpha)$ quantile of the posterior distribution of $S$. We simply use 
the $(1-\alpha)$ sample quantile, denoted as $\hat{\lambda}_{\text{BSCB}}$, of 
a large number $L$ of i.i.d. samples of $S$, $\{S^{(1)},\dots,S^{(L)}\}$, to approximate $\lambda$. Since the sample quantile converges to the population quantile almost surely as $L \rightarrow \infty$ \citep{serfling2009approximation}, $\hat{\lambda}_{\text{BSCB}}$ can be as close to $\lambda$ as desired if
$L$ is sufficiently large.

To simulate one $S$, we first simulate one $\bm{\theta}$ from its posterior distribution.
For the normal-gamma conjugate prior and the independent Jeffreys prior, this posterior is the multivariate $t$ distribution given in \eqref{NG_theta_posterior} and \eqref{eq_Jeffreys_posterior}, respectively; for other non-conjugate priors, the draw is taken from the HMC samples.
The posterior mean $E(\bm{\theta}|\bm{Y})$ and covariance matrix $\text{Cov}(\bm{\theta}|\bm{Y})$ can be obtained analogously.
The suprema of $T(x)$ can be attained only possibly at $a$, $b$ and the stationary points of the function $T(x)$.
Since the absolute value function in the numerator is not differentiable at $\bm{x}^T(\bm{\theta}-E(\bm{\theta}|\bm{Y}))=0$, we consider two approaches to address this issue. The first approach partitions the domain into differentiable subintervals and employs the R built-in function $\textit{optimize}$ to locate the local maximum. The second method seeks to find the stationary points of $h(x)=[T(x)]^2$ and reduces the problem to identifying the real roots of a $(4p-2)$th-order polynomial; see Web Appendix \ref{Appendix_optimization} for details.

Finally, the BSCB is given by
\begin{equation}
    \mathcal{I}_{\text{BSCB}}=\bm{x}^TE(\bm{\theta}|\bm{Y}) \pm \hat{\lambda}_{\text{BSCB}}\sqrt{\bm{x}^T\text{Cov}(\bm{\theta}|\bm{Y})\bm{x}} \, .
    \label{eq:BSCB}
\end{equation}

\subsection{Implementation}\label{subsect-implementation}

We present the procedures for constructing the  $(1-\alpha)$ BSCB for $\bm{x}^T\bm{\theta}$ over $x\in(a,b)$ in Algorithm \ref{alg}, which is implemented in R and realized in an accompanying R package $\textit{BSCB}$. 
In Step $1$, for the conjugate prior and the independent Jeffreys prior methods, $\bm{\theta}^{(l)}$ is generated using the function $\textit{rmvt}$ from the R package $\textit{mvtnorm}$ \citep{genz2009computation}. 
For the non-conjugate prior, the HMC is implemented via the \textit{cmdstanr} R package.
We recommend mean-centering the covariates before fitting the model, as this improves numerical stability and reduces collinearity issues in polynomial regression \citep{Budescu01101980}. 

\begin{algorithm}
\caption{The construction of the BSCB for  $\bm{x}^T\bm{\theta}$ over $x \in (a,b)$}
\DontPrintSemicolon
\KwIn{Design matrix $\bm{X}$, response vector $\bm{Y}$, covariate domain $x\in(a,b)$, confidence level $1-\alpha$, Monte Carlo sample size $L$, and HMC sample size $N$ (if needed)}
\KwOut{Estimated critical constant $\hat{\lambda}_{\mathrm{BSCB}}$ and BSCB $\mathcal{I}_{\mathrm{BSCB}}$}

$\textbf{Step 1: Bayesian modelling}$

1.1 Derive the posterior distribution of $\bm{\theta}$ under the chosen prior.\;

1.2 Compute the posterior mean $E(\bm{\theta}\mid\bm{Y})$ and covariance matrix $\mathrm{Cov}(\bm{\theta}\mid\bm{Y})$. Use closed-form expressions for conjugate prior or independent Jeffreys prior; otherwise, estimate them from HMC draws $\{\bm{\theta}^{(i)}\}_{i=1}^N$ as described in Section~\ref{sub-sub-sect-HMC}.

1.3 Generate i.i.d. posterior samples $\{\bm{\theta}^{(l)}\}_{l=1}^L$. Use direct sampling from the posterior for the conjugate prior or independent Jeffreys prior; for a non-conjugate prior, draw $L$ samples with replacement from the previously generated HMC draws $\{\bm{\theta}^{(i)}\}_{i=1}^N$.

$\textbf{Step 2: Computation of supremum statistics}$

\For{$l=1,\ldots,L$}{
    Compute
    \[
    S^{(l)}
    =
    \sup_{x\in(a,b)}
    \frac{\left|\bm{x}^T\bigl(\bm{\theta}^{(l)}-E(\bm{\theta}\mid\bm{Y})\bigr)\right|}
    {\sqrt{\bm{x}^T\mathrm{Cov}(\bm{\theta}\mid\bm{Y})\bm{x}}}.
    \]
}

$\textbf{Step 3. Critical constant and band construction}$

3.1 Estimate $\hat{\lambda}_{\text{BSCB}}$ as the $(1-\alpha)$ sample quantile of $\{S^{(l)}\}_{l=1}^L$.

3.2 Construct the BSCB as
\[
\mathcal{I}_{\mathrm{BSCB}}
=
\bm{x}^T E(\bm{\theta}\mid\bm{Y})
\pm
\hat{\lambda}_{\mathrm{BSCB}}
\sqrt{\bm{x}^T\mathrm{Cov}(\bm{\theta}\mid\bm{Y})\bm{x}}.
\]

\Return{$\hat{\lambda}_{\mathrm{BSCB}}$ \text{and} $\mathcal{I}_{\mathrm{BSCB}}$}
\label{alg}
\end{algorithm}

\subsection{Comparison with the frequentist band and the pointwise band}\label{subsect-comparison}

To evaluate the performance of BSCBs, we consider four benchmark methods: the simulation-based frequentist simultaneous confidence band (FSCB-S) of \citet{liu2008statistical},
the exact frequentist simultaneous confidence band (FSCB-E) of \citet{Liu2014}, the frequentist pointwise confidence band (FPCB), and the Bayesian pointwise credible band (BPCB).

The FSCBs have the same form,
\[
\mathcal{I}_{\text{FSCB}} = \bm{x}^T\hat{\bm{\theta}}_{\text{GLS}} \pm\hat{\lambda}_{\text{FSCB}}\hat{\sigma}_{\text{GLS}}\sqrt{\bm{x}^T(\bm{X}^TV^{-1}\bm{X})^{-1}\bm{x}},
\]
with the two methods differing only in the critical constant  $\hat{\lambda}_{\text{FSCB}}$.
FSCB-E employs a numerical quadrature method that yields exact simultaneous coverage, but it can be computationally demanding for high-order polynomial models. 
In contrast, FSCB-S estimates $\hat{\lambda}_{\text{FSCB}}$ as the $(1-\alpha)$ quantile of the pivotal quantity $T=\sup_{x\in(a,b)}|\bm{x}^TW|/\sqrt{\bm{x}^T(\bm{X}^TV^{-1}\bm{X})^{-1}\bm{x}}$, where $ W\sim t_{n-p-1}(0,(\bm{X}^TV^{-1}\bm{X})^{-1})$.
Notably, under an independent Jeffreys prior, the BSCB-I-J is the same as the FSCB-S for any finite sample sizes; see the proof in Section~\ref{subsect_property_BSCB-I}.

To highlight the necessity of simultaneous inference, we also compare with two pointwise bands, which are constructed by connecting the confidence/credible interval at each point $x$.
The FPCB is given by:
\[
\mathcal{I}_{\text{FPCB}} = \bm{x}^T\hat{\bm{\theta}}_{\text{GLS}} \pm t_{n-p-1}^{\alpha/2}\hat{\sigma}_{\text{GLS}}\sqrt{\bm{x}^T(\bm{X}^TV^{-1}\bm{X})^{-1}\bm{x}},
\]
where $t_{n-p-1}^{\alpha/2}$ is the upper $\alpha/2$ quantile of the $t$ distribution with $n-p-1$ degrees of freedom.
Its Bayesian counterpart, the BPCB, is
\[
\mathcal{I}_{\text{BPCB}}= \bm{x}^TE(\bm{\theta}|\bm{Y}) \pm t_{\nu}^{\alpha/2}\sqrt{\bm{x}^T\text{Cov}(\bm{\theta}|\bm{Y})\bm{x}},
\]
when the marginal posterior distribution of $\bm{\theta}$ is available in closed form, with $E(\bm{\theta}|\bm{Y}),\text{Cov}(\bm{\theta}|\bm{Y})$ and $\nu$ given by \eqref{NG_theta_posterior} under the conjugate prior (BPCB-C) and by \eqref{eq_Jeffreys_posterior} under the independent Jeffreys prior (BPCB-I-J). 
In these cases, BPCB-I-J is the same as FPCB.
For non-conjugate priors implemented via HMC, the BPCB is constructed from the posterior samples as
\[
\mathcal{I}_{\text{BPCB-H}} = \left [ Q_{\alpha/2}(\bm{x}^T\bm{\theta}^{(i)}), Q_{1-\alpha/2}(\bm{x}^T\bm{\theta}^{(i)})\right],\quad i=1,\dots,N,
\]
where $Q_{\alpha/2}$ denotes the empirical $\alpha/2$-quantile based on the $N$ posterior draws.
Unlike simultaneous bands, FPCB and BPCB only satisfy $(1-\alpha)$ coverage for a fixed $x$, leading to a consistently lower simultaneous coverage over the interval $(a,b)$.

\section{Theoretical properties}\label{sect-theory}

In this section, we investigate the frequentist properties of the proposed BSCB.
We first prove that the general BSCB is asymptotically equivalent to the FSCB, attaining the nominal $1-\alpha$ frequentist coverage as $n\to\infty$ under mild regularity conditions.
We then establish the finite-sample property of the BSCB-I-J, showing that it is the same as the FSCB.

\subsection{Asymptotic property of the general BSCB}\label{subsect_Freq_Coverage_Proof}

We establish that the proposed BSCB for the regression curve $\bm{x}^T\bm{\theta}$ over $x\in(a,b)$ attains the nominal frequentist simultaneous coverage level $1-\alpha$ as $n\to\infty$, i.e.,
\[
P_{\bm{\theta}_0}\left\{ \bm{x}^T\bm{\theta}_0 \in \mathcal{I}_{\text{BSCB}}(x)\ \ \forall x \in (a,b)\right\}\to 1-\alpha.
\]
The argument relies on the Bernstein-von Mises theorem \citep{Vaart_1998}, under standard regularity conditions of model identifiability, local smoothness of the log-likelihood, and a non-singular Fisher information matrix; see Web Appendix \ref{Appendix_Th1}.

\begin{theorem}\label{Th1}
Under the regularity conditions detailed in Web Appendix \ref{Appendix_Th1}, the proposed BSCB is asymptotically equivalent to the FSCB  as the sample size $n\to \infty$.
Consequently, the BSCB attains the nominal frequentist coverage rate of $1-\alpha$ asymptotically, that is, 
\[P_{\bm{\theta}_0}\left\{ \bm{x}^T\bm{\theta}_0 \in \mathcal{I}_{\text{BSCB}}\quad \forall x \in (a,b)\right\}\to 1-\alpha \quad \text{as} \quad n \to \infty.\]
\end{theorem}
Theorem \ref{Th1} establishes an asymptotic link between the general BSCB under a broad class of priors and the classical FSCB.
To prove the theorem, it suffices to show the asymptotic equivalence of (i) the point estimator, (ii) the bandwidth, and (iii) the critical constant, the details of which are provided in Appendix \ref{Appendix_Th1}.

\subsection{Finite-sample property of BSCB-I-J}\label{subsect_property_BSCB-I}

The theorem below establishes that the BSCB-I-J is the same as the FSCB even though they have different interpretations.

\begin{theorem}\label{Th2}
Under the independent Jeffreys prior, the BSCB-I-J is identical to the FSCB for any finite sample size $n>p+1$.
\end{theorem}
Theorem \ref{Th2} demonstrates that by using the independent Jeffreys prior, even in a small-sample setting, BSCB is exactly the same as FSCB, thus attaining the same nominal frequentist coverage; see detailed proof in Web Appendix \ref{Appendix_Th2}.
Therefore, the BSCB-I-J could serve as an objective baseline for further analysis, and any differences from other BSCB methods can be attributed to the influence of the prior through the resulting posterior distribution and then the critical constant.

\section{Simulations}\label{sect-simulation}
The effectiveness of the proposed BSCB is assessed through two simulation studies featuring quadratic and cubic models under i.i.d. and autoregressive error assumptions.

\subsection{Simulation settings}
\subsubsection{Example 1: quadratic and cubic model with i.i.d. errors}
In Example 1, we consider the quadratic model $(p=2)$ and the cubic model $(p=3)$ defined by:
\[
Y_i=\sum_{j=0}^p\theta_j x_i^j+e_i,\quad i=1,\dots,n.
\]
To obtain a precise estimation, we employ a D-optimal (DO) design \citep{kiefer1959optimum} to determine the design points $x_i$ within the interval $[-5,5]$.
The design matrix $\bm{X}$ remains fixed across $K=1000$ replications.
The random error terms $e_i$ are independently drawn from $N(0,\sigma^2V)$, with $\sigma\in\{0.1,0.2,0.5,1\}$. 
$V$ is assumed to be the identity matrix to give i.i.d. errors.
We consider sample sizes of $n\in\{20,50,100,200\}$.
Following \citet{trout1973uniform} 
and Example 3.16 in \citet{Fahrmeir2021}, the parameter vectors are set to be $\bm{\theta} = (-6, -3, 0.25)^{T}$ for the quadratic case and $\bm{\theta} = (1, 2, -1, 0.5)^{T}$ for the cubic case.

\subsubsection{Example 2: quadratic model with autoregressive errors}\label{subsect:simulation_AR}

Example 2 evaluates a quadratic model with an AR(1) error structure:
\[
Y_i = \theta_0+\theta_1x_{i}+\theta_2x_i^2+\varepsilon_i, \quad i=1,\dots,n
\]
In this setup, the covariates $x_i$, sample sizes $n$, and values of the true parameter $\bm{\theta}$ are the same as in Example 1.
The error terms satisfy $\varepsilon\sim N(0,\sigma^2V)$, with $\sigma \in \{0.2, 0.5, 1\}$.
The covariance matrix $V$ has entries $v_{ij}=\rho^{|i-j|}/(1-\rho^2)$, where the autoregressive parameter $\rho \in \{0.2,0.5,0.8\}$ is assumed known. 
This setting is particularly relevant to longitudinal clinical trials, in which patients receive a fixed dose and are repeatedly assessed over time. 
The BSCB allows us to check whether treatment effects are sustained over time.

\subsection{Evaluation criteria}
To provide a comprehensive assessment from both Bayesian and frequentist perspectives, we consider two evaluation criteria with different emphases: 
the posterior simultaneous coverage probability (PSCP) and the empirical simultaneous coverage rate (ESCR).

\subsubsection{PSCP}

PSCP is defined as the posterior probability that the Bayesian band $\mathcal{I}$ covers the random regression function
over the covariate range $(a,b)$, conditional on the observed data $\bm{Y}$,
\begin{equation}
    \text{PSCP} = P_{\bm{\theta|\bm{Y}}} \{ \bm{x}^T\bm{\theta} \in \mathcal{I} \quad \forall x\in(a,b)  \}.
    \label{def_PSCP}
\end{equation}
By construction, the PSCP of the BSCB should be exactly equal to the nominal level $1-\alpha$, whereas that of the BPCB is generally smaller; see Appendix~\ref{appendix:PSCP}.
For the BSCB, we can further rewrite \eqref{def_PSCP} as:
\[
\begin{aligned}
    &P_{\bm{\theta}|\bm{Y}} \{ \bm{x}^T\bm{\theta} \in \mathcal{I}_{\text{BSCB}} \quad \forall x\in(a,b)  \}\nonumber \\
     = & P_{\bm{\theta}|\bm{Y}} \left\{ \sup_{x\in(a,b)}\frac{|\bm{x}^T(\bm{\theta}-E(\bm{\theta}|\bm{Y}))|}{\sqrt{\bm{x}^T\text{Cov}(\bm{\theta}|\bm{Y})\bm{x}}} \le \hat{\lambda}_{\text{BSCB}} \right \} \nonumber \\
     = & P_{\bm{\theta}|\bm{Y}} \left\{ \sup_{x\in(a,b)} T(x) \le \hat{\lambda}_{\text{BSCB}} \right\} .
\end{aligned}
\]
This PSCP can be estimated by Monte Carlo simulation using $G$ samples $\{\bm{\theta}^{(j)}\}_{j=1}^G$ drawn from the posterior distribution, as in \citep{besag1995bayesian,held2004simultaneous}.
Specifically, let
\[
T_{\sup}^{(j)} = \sup_{x \in (a,b)} T^{(j)}(x) = \sup_{x \in (a,b)} \frac{|\bm{x}^T(\bm{\theta}^{(j)}-E(\bm{\theta}|\bm{Y}))|}{\sqrt{\bm{x}^T\text{Cov}(\bm{\theta}|\bm{Y})\bm{x}}}.
\]
Then the PSCP estimate is given by 
\[
\widehat{\text{PSCP}} = \frac{1}{G}\sum_{j=1}^G\textbf{1}\{T_{\sup}^{(j)}\le \hat{\lambda}\},
\]
which converges to the true value with probability 1 as $G\to \infty$. $\widehat{\text{PSCP}}$ measures band performance for one realization of the data.
To rigorously evaluate the robustness of the method, we report the mean posterior coverage probability (MPSCP) over $K$ independently simulated datasets: 
\[
\text{MPSCP}=\frac{1}{K}\sum_{k=1}^K \widehat{\text{PSCP}}^{(k)}.
\]

\subsubsection{ESCR}

Although BSCB is developed within a Bayesian framework, we further examine its frequentist coverage property by ESCR, which is the most commonly used criterion in the existing frequentist literature \citep{Naiman1986, Liu2014,  Lin01022016}. 
It is defined as the proportion of the true regression function $\bm{x}^T\bm{\theta}_0$ that is entirely contained within the band on $x\in(a,b)$ over $K$ Monte Carlo replications.
In each replication, the design matrix $\bm{X}$ is kept fixed, whereas the response vector $\bm{Y}$ varies due to the generated error term.
Let $\mathcal{I}_i=[L_i(x), U_i(x)]$ be the band in the $i$-th replication, where $L_i(x)$ is the lower bound function and $U_i(x)$ is the upper bound function of the band.
Then $\text{ESCR}
=
\frac{1}{K}\sum_{i=1}^K
\mathbf{1}\{L_i(x)\le \bm{x}^T\bm{\theta}_0 \le U_i(x)\ \forall x\in(a,b)\}$, where $\bm{\theta}_0$ is the true parameter under the frequentist approach.
Implementation details are provided in the Web Appendix \ref{appendix:ESCR}.

Notably, the BSCB is not constructed to guarantee the frequentist coverage.
Compared to the FSCB, the BSCB is expected to be narrower due to the incorporation of prior information, which may result in an ESCR slightly below the nominal level $1-\alpha$ in small-sample scenarios. 
Nevertheless, when the prior information is well calibrated, the BSCB can provide more efficient inference by yielding tighter bands while maintaining ESCR close to $1-\alpha$.

\subsection{Computational setup}\label{subsecct-comp}

Following Section~\ref{subsect-posterior}, we evaluate six types of BSCBs at the nominal confidence level $1-\alpha=0.95$. 
Table~\ref{tab:hyperparameters_priors} summarizes their prior distributions and corresponding hyperparameter settings; see the detailed discussion in Web Appendix~\ref{Appendix_Simulation_Hyperparameters}.
For the normal-gamma conjugate prior, we consider the hyperparameters elicited by the empirical Bayes method (BSCB-C-E), the unit-information method (BSCB-C-U), and the g-prior (BSCB-C-G).
For non-conjugate priors implemented via HMC, we employ the normal-half-normal prior (BSCB-H-N) and the normal-half-Cauchy prior (BSCB-H-C). 
Additionally, the independent Jeffreys prior (BSCB-I-J) is included as a non-informative case, the resulting band of which is the same as FSCB-S (and FSCB-E when the number of replications is sufficiently large).
To provide a pointwise benchmark, we also include BPCB-I-J, which is equivalent to FPCB.
For each BSCB method, the critical value $\hat{\lambda}_{\text{BSCB}}$ is estimated by Monte Carlo simulation with $L=500,000$ samples.
All computations were performed on a MacBook Pro (Apple M1 Chip with 8-core CPU, 8 GB RAM).

\begin{table}
\centering
\caption{Summary of prior distributions and hyperparameter settings for the simulation studies. }
\resizebox{\textwidth}{!}{%
\begin{tabular}{llllllll}
\toprule
\multicolumn{2}{c}{Method}
    & \multicolumn{2}{c}{Prior Distribution}
    & \multicolumn{4}{c}{Hyperparameters} \\
\midrule
\multirow{3}{*}{\shortstack[l]{Conjugate}}
    & BSCB-C-E
    & \multirow{3}{*}{$\bm{\theta}\mid\tau \sim N(\bm{\mu},(\tau\mathcal{P})^{-1})$}
    & \multirow{3}{*}{$\tau \sim \text{Gamma}(\alpha_0,\beta_0)$}
    & $\bm{\mu}=\hat{\bm{\theta}}_{\text{GLS}}$
    & $\mathcal{P}=10^{-3}I_{p+1}$
    & $\alpha_0=1$
    & $\beta_0=\hat{\sigma}_{\text{GLS}}^2$ \\
    & BSCB-C-U & &
    & $\bm{\mu}=\hat{\bm{\theta}}_{\text{GLS}}$
    & $\mathcal{P}=\tfrac{1}{n}\bm{X}^TV^{-1}\bm{X}$
    & $\alpha_0=\tfrac{1}{2}$
    & $\beta_0=\tfrac{1}{2}\hat{\sigma}_{\text{GLS}}^2$ \\
    & BSCB-C-G & &
    & $\bm{\mu}=\hat{\bm{\theta}}_{\text{GLS}}$
    & $\mathcal{P}=\tfrac{1}{g}\bm{X}^TV^{-1}\bm{X}$
    & $\alpha_0=\tfrac{1}{2}$
    & $\beta_0=\tfrac{1}{2}\hat{\sigma}_{\text{GLS}}^2$ \\
\midrule
\multirow{2}{*}{\shortstack[l]{Non-conjugate\\via HMC}}
    & BSCB-H-N
    & $\bm{\theta} \sim N(\bm{\mu},\sigma_{\bm{\theta}}^2I_{p+1})$
    & $\sigma \sim \text{half-Normal}(0,5^2)$
    & $\bm{\mu}=\bm{0}$
    & $\sigma_{\bm{\theta}}=10$
    & ---
    & --- \\
    & BSCB-H-C
    & $\bm{\theta} \sim N(\bm{\mu},\sigma_{\bm{\theta}}^2I_{p+1})$
    & $\sigma \sim \text{half-Cauchy}(0,2)$
    & $\bm{\mu}=\hat{\bm{\theta}}_{\text{GLS}}$
    & $\sigma_{\bm{\theta}}=\hat{\sigma}_{\text{GLS}}$
    & ---
    & --- \\
\midrule
\shortstack[l]{Jeffreys}
    & BSCB-I-J
    & $\bm{\theta}\mid\tau \propto 1$
    & $\tau \propto \tau^{-1}$
    & ---
    & ---
    & ---
    & --- \\
\bottomrule
\end{tabular}}\\[0.5em]
\parbox{\textwidth}{\small \textit{Notes:} The BSCB methods are categorized by the chosen priors: (1) normal--gamma conjugate priors with hyperparameters elicited by the empirical Bayes method (BSCB-C-E), unit-information prior (BSCB-C-U), and $g$-prior (BSCB-C-G); (2) non-conjugate priors implemented via HMC: the normal--half-normal prior (BSCB-H-N) and normal--half-Cauchy prior (BSCB-H-C); and (3) the independent Jeffreys prior (BSCB-I-J).}
\label{tab:hyperparameters_priors}
\end{table}

Sensitivity analyses for the conjugate prior suggest that posterior inference is robust to the chosen hyperparameter values, as reported in Web Appendix~\ref{Appendix_Simulation_Sensitivity}. 
Prior predictive checks and posterior predictive checks for all priors in Web Appendix~\ref{Appendix_Simulation_PPC} further support that these priors are reasonable and that the resulting models fit the data adequately.
For the HMC method, the Gelman--Rubin diagnostic $\hat{R}$, effective sample size (ESS), and the autocorrelation plot in Web Appendix~\ref{Appendix_Simulation_HMC}, indicate satisfactory convergence and mixing of the HMC chains.

\subsection{Simulation results}

\begin{figure}
\centerline{
\includegraphics[width=1.0\linewidth]{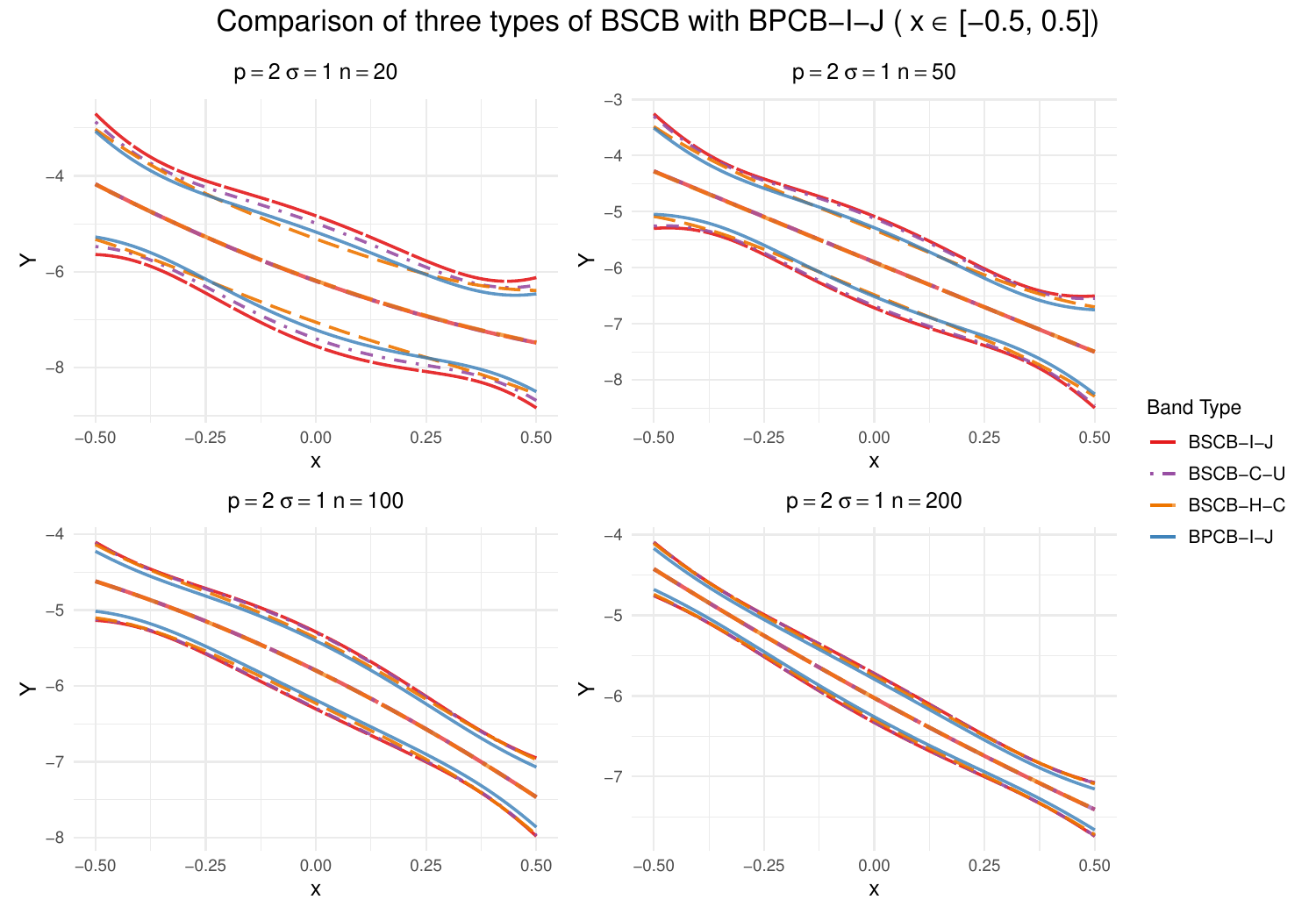}}
\caption{Comparison of three types of BSCB with BPCB-I-J for the quadratic model $(p=2,\sigma=1)$.
BSCB-I-J: BSCB under the independent Jeffreys prior; BSCB-C-U: BSCB under the normal-gamma conjugate prior with the unit information hyperparameters; BSCB-H-C: BSCB under the normal-half Cauchy (0,2) prior implemented via HMC; BPCB-I-J: BPCB under the independent Jeffreys prior, included as a Bayesian pointwise reference. Results are shown for $n=20,50,100,200$.}
\label{fig:3BSCB_BPCB_p2_sigma1}
\end{figure}

\begin{figure}
\centerline{
\includegraphics[width=1.0\linewidth]{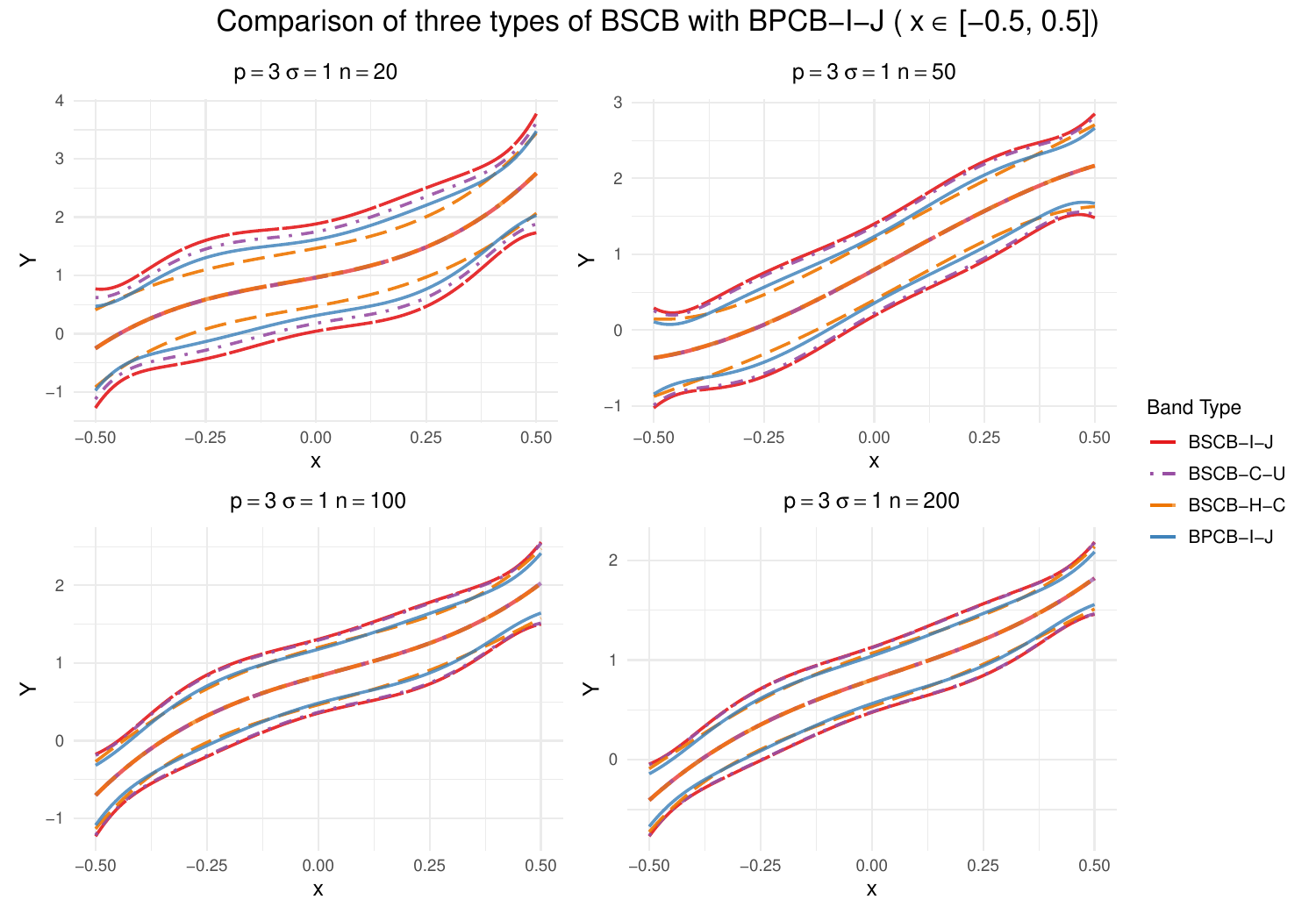}}
\caption{Comparison of three types of BSCB with BPCB-I-J for the cubic model $(p=3,\sigma=1)$.
BSCB-I-J: BSCB under the independent Jeffreys prior; BSCB-C-U: BSCB under the normal-gamma conjugate prior with the unit information hyperparameters; BSCB-H-C: BSCB under the normal-half Cauchy (0,2) prior implemented via HMC; BPCB-I-J: BPCB under the independent Jeffreys prior, included as a Bayesian pointwise reference. Results are shown for $n=20,50,100,200$.}
\label{fig:3BSCB_BPCB_p3_sigma1}
\end{figure}

To illustrate differences among the BSCBs, we provide an additional example over a narrow interval $x \in [-0.5, 0.5]$ under the same settings as Example 1. 
Figures~\ref{fig:3BSCB_BPCB_p2_sigma1} and \ref{fig:3BSCB_BPCB_p3_sigma1} focus on selected comparisons between BSCB-C-U, BSCB-H-C, BSCB-I-J, and BPCB-I-J for the quadratic and cubic models, with more comprehensive results provided in Web Appendix~\ref{Appendix_simulation_plots}.
This includes comparisons of conjugate priors (Web Figures~\ref{fig:3BSCB_Conjugate_p2_sigma1}-\ref{fig:3BSCB_Conjugate_p3_sigma1}), non-conjugate priors (Web Figures~\ref{fig:3BSCB_HMC_p2_sigma1}-\ref{fig:3BSCB_HMC_p3_sigma1}), and the frequentist methods (Web Figures~\ref{fig:All_p2_sigma1}-\ref{fig:All_p3_sigma1}).

Across all scenarios, three general patterns emerge as expected. 
First, the incorporation of prior information leads to narrower BSCBs compared to the objective BSCB-I-J, or equivalently, the FSCB.
These differences in bandwidth are most pronounced when the sample size is limited ($n=20$).
Based on the comprehensive simulation results, the BSCB methods can be ranked by their bandwidth from widest to narrowest as follows: $\text{BSCB-I-J} \approx \text{BSCB-H-N} > \text{BSCB-C-E} > \text{BSCB-C-G} \approx \text{BSCB-C-U}  > \text{BSCB-H-C}.$
This suggests that the normal-half-Cauchy prior and the conjugate prior are more informative compared to the normal-half-normal prior.
Second, simultaneous bands remain consistently wider than the corresponding pointwise bands within the same framework (e.g., BSCB is wider than BPCB). However, a highly informative BSCB, such as BSCB-H-C, can occasionally be narrower than a non-informative BPCB (BPCB-I-J), as illustrated in Web Figures \ref{fig:3BSCB_HMC_p2_sigma1} and \ref{fig:3BSCB_HMC_p3_sigma1}.
Third, as $n$ increases, all BSCBs shrink and the influence of the prior diminishes.

Despite these differences in bandwidth, the choice of prior has a negligible impact on point estimation. 
The posterior mean estimates for all methods align closely with the frequentist GLS curve, indicating that the priors primarily affect the posterior variance rather than the estimation of $\bm{\theta}$.
This is further supported by the boxplots of the posterior samples in Web Figures~\ref{fig:theta_boxplot_2_1_20_comparison_4}--\ref{fig:theta_boxplot_3_1_20_comparison_4}.

Table~\ref{tab_MPSCP_IID} summarizes the MPSCP of the three BSCBs and  BPCB-I-J among $K=1000$ replications with $G=10{,}000$ Monte Carlo samples per replication. 
All BSCB methods attain the nominal level exactly, whereas the pointwise method exhibits undercoverage as expected. 

Table~\ref{tab_ESCR_IID} presents the corresponding ESCR results.
All BSCB methods achieve ESCR close to the nominal level, consistent with the theoretical arguments in Section \ref{sect-theory}. 
Minor deviations from  $1-\alpha$ are due to sampling variation and diminish as the number of replications $K$ increases; see Web Table \ref{tab:replication} in Web Appendix~\ref{Appendix_simulation_tables}.
All methods are computationally feasible, with average runtimes below 0.8 minutes for BSCB-C and BSCB-I-J and below 2.8 minutes for BSCB-H under 6-core parallel computing, as shown in Web Table \ref{tab:time}.

\begin{table}
\centering
\caption{Mean posterior simultaneous coverage probability of the 95\% BSCB-H-N, BSCB-H-C, BSCB-C-E, BSCB-C-U, BSCB-I-J and BPCB-I-J for the quadratic model and the cubic model with i.i.d. random errors.}
\label{tab_MPSCP_IID}
\begin{tabular}{ccccccccc}
\toprule
$p$ & $\sigma$ & $n$ & \makecell{BSCB-H-N} & \makecell{BSCB-H-C} & \makecell{BSCB-C-E} & \makecell{BSCB-C-U} & \makecell{BSCB-I-J} & \makecell{BPCB-I-J} \\
\midrule
2 & 0.1 & 20  & 0.950 & 0.950 & 0.950 & 0.950 & 0.950 & 0.821 \\
  &     & 50  & 0.950 & 0.950 & 0.950 & 0.950 & 0.950 & 0.813 \\
  &     & 100 & 0.950 & 0.950 & 0.950 & 0.950 & 0.950 & 0.811 \\
  &     & 200 & 0.950 & 0.950 & 0.950 & 0.950 & 0.950 & 0.810 \\
  & 0.2 & 20  & 0.950 & 0.950 & 0.950 & 0.950 & 0.950 & 0.821 \\
  &     & 50  & 0.950 & 0.950 & 0.950 & 0.950 & 0.950 & 0.813 \\
  &     & 100 & 0.950 & 0.950 & 0.950 & 0.950 & 0.950 & 0.811 \\
  &     & 200 & 0.950 & 0.950 & 0.950 & 0.950 & 0.950 & 0.809 \\
  & 0.5 & 20  & 0.950 & 0.950 & 0.950 & 0.950 & 0.950 & 0.821 \\
  &     & 50  & 0.950 & 0.950 & 0.950 & 0.950 & 0.950 & 0.813 \\
  &     & 100 & 0.950 & 0.950 & 0.950 & 0.950 & 0.950 & 0.811 \\
  &     & 200 & 0.950 & 0.950 & 0.950 & 0.950 & 0.950 & 0.809 \\
  & 1   & 20  & 0.950 & 0.950 & 0.950 & 0.950 & 0.950 & 0.821 \\
  &     & 50  & 0.950 & 0.950 & 0.950 & 0.950 & 0.950 & 0.813 \\
  &     & 100 & 0.950 & 0.950 & 0.950 & 0.950 & 0.950 & 0.811 \\
  &     & 200 & 0.950 & 0.950 & 0.950 & 0.950 & 0.950 & 0.810 \\
\midrule
3 & 0.1 & 20  & 0.950 & 0.950 & 0.950 & 0.950 & 0.950 & 0.767 \\
  &     & 50  & 0.950 & 0.950 & 0.950 & 0.950 & 0.950 & 0.753 \\
  &     & 100 & 0.950 & 0.950 & 0.950 & 0.950 & 0.950 & 0.749 \\
  &     & 200 & 0.950 & 0.950 & 0.950 & 0.950 & 0.950 & 0.747 \\
  & 0.2 & 20  & 0.950 & 0.950 & 0.950 & 0.950 & 0.950 & 0.767 \\
  &     & 50  & 0.950 & 0.950 & 0.950 & 0.950 & 0.950 & 0.753 \\
  &     & 100 & 0.950 & 0.950 & 0.950 & 0.950 & 0.950 & 0.734 \\
  &     & 200 & 0.950 & 0.950 & 0.950 & 0.950 & 0.950 & 0.744 \\
  & 0.5 & 20  & 0.950 & 0.950 & 0.950 & 0.950 & 0.950 & 0.767 \\
  &     & 50  & 0.950 & 0.950 & 0.950 & 0.950 & 0.950 & 0.753 \\
  &     & 100 & 0.950 & 0.950 & 0.950 & 0.950 & 0.950 & 0.749 \\
  &     & 200 & 0.950 & 0.950 & 0.950 & 0.950 & 0.950 & 0.743 \\
  & 1   & 20  & 0.950 & 0.950 & 0.950 & 0.950 & 0.950 & 0.767 \\
  &     & 50  & 0.950 & 0.950 & 0.950 & 0.950 & 0.950 & 0.753 \\
  &     & 100 & 0.950 & 0.950 & 0.950 & 0.950 & 0.950 & 0.749 \\
  &     & 200 & 0.950 & 0.950 & 0.950 & 0.950 & 0.950 & 0.746 \\
\bottomrule
\end{tabular}
\end{table}

\begin{table}
\centering
\caption{Empirical simultaneous coverage rate of the 95\% BSCB-H-N, BSCB-H-C, BSCB-C-E, BSCB-C-U, BSCB-I-J, and BPCB-I-J for the quadratic model and the cubic model with i.i.d.\ random errors.}
\label{tab_ESCR_IID}
\begin{tabular}{ccccccccc}
\toprule
$p$ & $\sigma$ & $n$ & \makecell{BSCB-H-N} & \makecell{BSCB-H-C} & \makecell{BSCB-C-E} & \makecell{BSCB-C-U} & \makecell{BSCB-I-J} & \makecell{BPCB-I-J} \\
\midrule
2 & 0.1 & 20  & 0.963 & 0.952 & 0.930 & 0.920 & 0.958 & 0.810 \\
  &     & 50  & 0.945 & 0.941 & 0.929 & 0.927 & 0.939 & 0.800 \\
  &     & 100 & 0.950 & 0.948 & 0.945 & 0.944 & 0.948 & 0.818 \\
  &     & 200 & 0.949 & 0.948 & 0.946 & 0.946 & 0.949 & 0.817 \\
  & 0.2 & 20  & 0.959 & 0.951 & 0.913 & 0.904 & 0.950 & 0.814 \\
  &     & 50  & 0.939 & 0.940 & 0.929 & 0.927 & 0.939 & 0.800 \\
  &     & 100 & 0.950 & 0.951 & 0.947 & 0.946 & 0.953 & 0.810 \\
  &     & 200 & 0.946 & 0.945 & 0.945 & 0.945 & 0.946 & 0.819 \\
  & 0.5 & 20  & 0.959 & 0.949 & 0.913 & 0.904 & 0.950 & 0.814 \\
  &     & 50  & 0.944 & 0.938 & 0.929 & 0.927 & 0.939 & 0.800 \\
  &     & 100 & 0.950 & 0.948 & 0.947 & 0.946 & 0.953 & 0.810 \\
  &     & 200 & 0.949 & 0.943 & 0.945 & 0.944 & 0.946 & 0.810 \\
  & 1   & 20  & 0.960 & 0.946 & 0.913 & 0.904 & 0.950 & 0.814 \\
  &     & 50  & 0.942 & 0.939 & 0.929 & 0.927 & 0.939 & 0.800 \\
  &     & 100 & 0.954 & 0.951 & 0.949 & 0.946 & 0.952 & 0.811 \\
  &     & 200 & 0.947 & 0.945 & 0.945 & 0.945 & 0.947 & 0.816 \\
\midrule
3 & 0.1 & 20  & 0.963 & 0.951 & 0.907 & 0.892 & 0.953 & 0.776 \\
  &     & 50  & 0.950 & 0.950 & 0.936 & 0.930 & 0.949 & 0.744 \\
  &     & 100 & 0.947 & 0.941 & 0.933 & 0.929 & 0.946 & 0.709 \\
  &     & 200 & 0.949 & 0.950 & 0.948 & 0.948 & 0.950 & 0.741 \\
  & 0.2 & 20  & 0.963 & 0.954 & 0.907 & 0.892 & 0.953 & 0.776 \\
  &     & 50  & 0.953 & 0.949 & 0.936 & 0.929 & 0.949 & 0.744 \\
  &     & 100 & 0.948 & 0.945 & 0.941 & 0.938 & 0.947 & 0.713 \\
  &     & 200 & 0.954 & 0.954 & 0.950 & 0.950 & 0.953 & 0.738 \\
  & 0.5 & 20  & 0.964 & 0.950 & 0.907 & 0.892 & 0.953 & 0.776 \\
  &     & 50  & 0.958 & 0.956 & 0.936 & 0.931 & 0.954 & 0.733 \\
  &     & 100 & 0.949 & 0.941 & 0.933 & 0.929 & 0.946 & 0.709 \\
  &     & 200 & 0.950 & 0.953 & 0.948 & 0.947 & 0.949 & 0.734 \\
  & 1   & 20  & 0.964 & 0.948 & 0.907 & 0.892 & 0.953 & 0.776 \\
  &     & 50  & 0.952 & 0.951 & 0.936 & 0.929 & 0.949 & 0.744 \\
  &     & 100 & 0.941 & 0.945 & 0.933 & 0.929 & 0.946 & 0.709 \\
  &     & 200 & 0.949 & 0.950 & 0.951 & 0.950 & 0.952 & 0.742 \\
\bottomrule
\end{tabular}
\end{table}

Web Table~\ref{tab_MPSCP_AR} and Web Table~\ref{tab_ESCR_AR} present the MPSCP and ESCR in Example 2.
The results under the AR(1) error structure are broadly consistent with those obtained in the i.i.d. setting, indicating BSCB is robust not only to the noise level and sample size but also to autoregressive dependence in the errors.

\subsection{Practical recommendations for choosing among BSCB methods}

To facilitate the practical implementation of BSCB, we offer the following recommendations. The conjugate prior (BSCB-C) is preferred when reliable expert knowledge supports its use. 
Among the weakly informative hyperparameter choices, BSCB-C-U and BSCB-C-G are more informative than BSCB-C-E. 
When the conjugate prior may be unsuitable (e.g., if $\sigma$ is close to zero), the non-conjugate prior via the HMC method (BSCB-H) is recommended for its flexibility and robustness, with BSCB-H-C imposing stronger shrinkage than BSCB-H-N.
Finally, in the absence of prior information, the independent Jeffreys prior (BSCB-I-J) should be employed as an objective baseline that connects naturally with frequentist methods.

\section{Application to minimum effective dose estimation in a Phase II dose-response study}\label{sect-realexample}
Finding the dose-response relationship and identifying the optimal dose is a key objective throughout the drug development process. 
In this study, we analyze a dataset from a Phase II randomized double-blind parallel group trial, originally discussed by \citet{bretz2005combining} and \citet{Liu2014}. 
The trial involved 100 patients, randomly assigned equally to receive a placebo or one of four active doses ($x=0.05, 0.2, 0.6, 1$), with 20 patients per group.
For confidentiality, the actual doses were scaled to $[0,1]$, where $x=a=0$ represents the placebo treatment.
The response variable $Y$ indicates the treatment efficacy and is assumed to be normally distributed. 
The observed mean responses and standard deviations for the five treatment groups are summarized in Web Table~\ref{table-clinical} of Web Appendix \ref{Appendix_Real_Data}.

According to \citet{Liu2014}, the best-fitting model is the quadratic model compared with the linear model and the cubic model, which is given by
\[
Y_i=\theta_0+\theta_1x_i+\theta_2 x_i^2+\varepsilon_i,\quad \varepsilon_i\sim N(0,\sigma^2).
\]
To quantify the possible range of the dose-response curve, we construct BSCB-H-N, BSCB-C-U, and BSCB-I-J over $[0,1]$.
As in Section~\ref{subsect-implementation}, the BSCBs are constructed on the mean-centered covariates and mapped back to the original covariate scale for interpretation.
This centering scheme does not affect the critical values of BSCB-C and BSCB-I-J, while improving the sampling efficiency and estimation accuracy of BSCB-H; see Web Table~\ref{table_real_mean_center}.

\begin{figure}
\centerline{%
\includegraphics[width=1.0\linewidth]{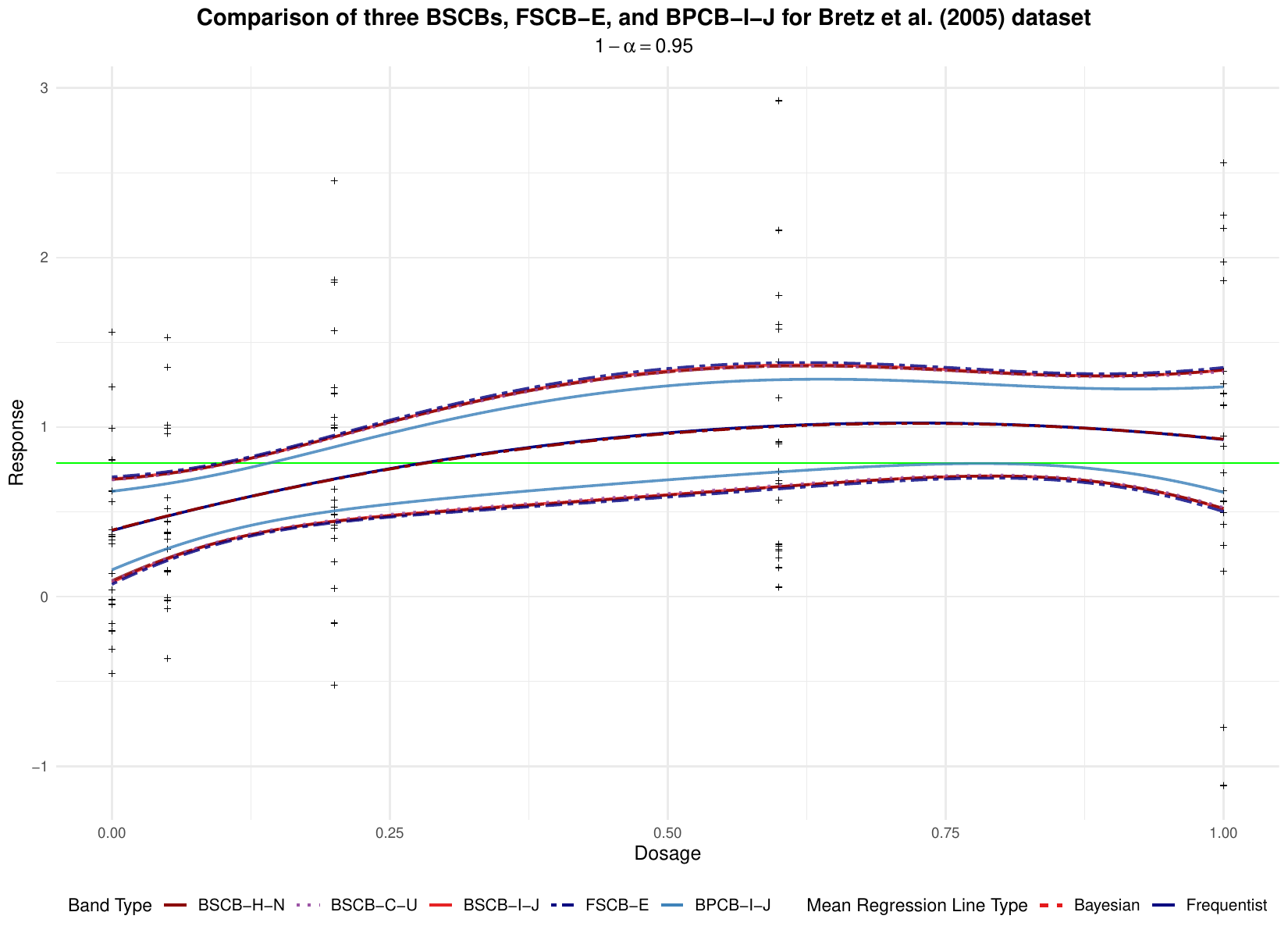}}
\caption{The 95\% BSCB-H-N, BSCB-C-U, BSCB-I-J, FSCB-E, and BPCB-I-J for the dataset of \citet{bretz2005combining}. 
The bands are constructed using mean-centered covariates and plotted on the original covariate scale. The horizontal green line indicates $Y = \bm{a}^T\bm{\theta} + \Delta$, which is 
used for MED estimation with clinical relevance threshold $\Delta = 0.4$.}
\label{fig_real2_centered_x}
\end{figure}

Figure~\ref{fig_real2_centered_x} compares three types of BSCB with FSCB-E from \citet{Liu2014} and the pointwise BPCB-I-J $(1-\alpha=0.95)$. 
The proposed BSCBs are nearly the same, a bit narrower than the FSCB-E, while BPCB-I-J lies on the inner side.

We next consider minimum effective dose (MED) estimation. The MED, defined as the smallest dose with a clinically relevant and statistically significant effect, is critical for guiding dose selection in subsequent Phase III confirmatory trials.
As a BSCB quantifies uncertainty simultaneously over the entire dose range, it provides multiplicity adjustment for MED estimation without conducting numerous hypothesis tests.
Following \citet{bretz2005combining}, the estimated MED can be determined by
\[
\widehat{MED} = \arg \min_{x\in[a,b]}\{U(\bm{x})>\bm{a}^T\bm{\theta}+\Delta, L(\bm{x})>\bm{a}^T\bm{\theta}\},
\]
where $\Delta$ is the clinical relevance threshold. 
Web Table~\ref{tab:smallest_dose} summarizes the resulting $\widehat{MED}$ estimates when $\Delta = 0.4$.
 All BSCB-based estimates lie between the FSCB-E estimate $(0.0994)$ and the Bayesian pointwise estimate BPCB-I-J $(0.1427)$. 
Among the BSCB methods,  BSCB-C-U yields a slightly larger estimate $(0.1094)$, followed by BSCB-H-N $(0.1087)$ and BSCB-I-J $(0.1060)$, suggesting that stronger prior information leads to a slightly larger $\widehat{MED}$ estimate.

As noted in \citet{bretz2025dose}, many drugs were initially marketed at doses that were later recognized as excessive, often positioned on the plateau of the dose-response curve for the desired effect. 
For instance, \citet{Sacks2014} reported that 32\% of FDA-approved drugs (2000--2012) required resubmission, often due to uncertainties in dose selection. 
Consequently, the MED derived from the pointwise methods may carry unacceptable safety risks, whereas the FSCB-based estimate may be too low to achieve adequate efficacy.
The BSCB framework offers a balanced alternative.
Specifically, BSCB-C produces a slightly larger estimate, whereas BSCB-H yields a smaller value, allowing researchers to tailor the dose selection strategy to different risk considerations.

\section{Discussion}\label{sect-conclusion}

In this study, we introduce a unified Bayesian framework for constructing two-sided hyperbolic Bayesian simultaneous credible bands for polynomial regression over a finite interval of the covariate. 
The proposed framework guarantees exact posterior simultaneous coverage, and is particularly useful for quantifying the uncertainty of the regression function under various assumptions about the prior knowledge.
To the best of our knowledge, this work appears to be the first to develop BSCBs for the regression function over a continuous covariate domain $(a,b)$ in a polynomial model.

The proposed methodology offers several advantages over existing methods.
First, the BSCBs provide exact posterior simultaneous coverage for all $x$ in the given interval, regardless of sample size, polynomial order, or error structure.
Compared with frequentist methods, BSCBs enable the incorporation of expert knowledge, thereby improving inference precision with narrower bands when informative priors are available. 
Furthermore, under an independent Jeffreys prior, the BSCB-I-J is identical to FSCB-S for finite samples.
We also demonstrate that as $n\to \infty$, the general BSCB is asymptotically equivalent to FSCB under mild regularity conditions, ensuring the frequentist coverage attains the nominal level.

From a practical perspective, BSCBs provide a direct tool for quantifying the efficacy uncertainty across the entire dose range.
As demonstrated in the Phase II clinical trial example, this approach supports decision-making toward the minimum effective dose, preserving the familywise error rate for infinitely many hypotheses simultaneously.

Although this study focuses on polynomial regression, the framework can be extended to the linear random-effects model in which the linear function is a polynomial.
In addition, it can be adapted to the Emax model, which is widely used for dose-response studies from a pharmacodynamic perspective. 
Furthermore, this methodology can be applied to quantile regression to construct BSCBs for percentile dose-response curves, facilitating the assessment of treatment efficacy across heterogeneous patient subgroups and supporting personalized medicine.
Overall, this study contributes a flexible and reliable approach to Bayesian simultaneous inference with broad applications in dose–response analysis and beyond.

\section*{Supplementary Material}
The supplementary material contains Web Appendices A-E. 

\newpage
\bibliography{reference}

\renewcommand{\figurename}{Web Figure}
\renewcommand{\tablename}{Web Table}
\SetAlgorithmName{Web Algorithm}{web algorithm}{List of Web Algorithms}
\renewcommand{\theequation}{S\arabic{equation}}
\setcounter{equation}{0}

\newcounter{webappdummy}
\renewcommand{\thewebappdummy}{\Alph{webappdummy}}

\newcounter{websubdummy}[webappdummy]
\renewcommand{\thewebsubdummy}{\thewebappdummy.\arabic{websubdummy}}

\newcounter{websubsubdummy}[websubdummy]
\renewcommand{\thewebsubsubdummy}{\thewebsubdummy.\arabic{websubsubdummy}}

\newcounter{webfigdummy}
\newcounter{webtabdummy}

\newcounter{webappendix}
\setcounter{webappendix}{0}
\renewcommand{\thewebappendix}{\Alph{webappendix}}

\newcommand{\webappendix}[1]{}  
\renewcommand{\webappendix}[1]{%  
  \refstepcounter{webappendix}%
  \setcounter{subsection}{0}%
  \phantomsection%
  \section*{Web Appendix \thewebappendix: #1}%
  \bookmark[dest=\HyperLocalCurrentHref, level=1]{Web Appendix \thewebappendix: #1}%
}
\renewcommand{\thesubsection}{\thewebappendix.\arabic{subsection}}

\input{dummy_labels}

\end{document}

%% file: dummy_labels.tex
% dummy_labels.tex
% Phantom labels for all \label{} in supplement.tex,
% preserving the A / B / B.1 / B.1.1 numbering hierarchy.
%
% -------------------------------------------------------
% PREAMBLE: add the following to the main .tex preamble
% -------------------------------------------------------
%
%   \newcounter{webappdummy}
%   \renewcommand{\thewebappdummy}{\Alph{webappdummy}}
%
%   \newcounter{websubdummy}[webappdummy]
%   \renewcommand{\thewebsubdummy}{\thewebappdummy.\arabic{websubdummy}}
%
%   \newcounter{websubsubdummy}[websubdummy]
%   \renewcommand{\thewebsubsubdummy}{\thewebsubdummy.\arabic{websubsubdummy}}
%
%   \newcounter{webfigdummy}
%   \newcounter{webtabdummy}
%
% -------------------------------------------------------
% USAGE: in main .tex, before \end{document}:
%   % \input{supplement}    <-- commented out
%   \input{dummy_labels}    <-- use this instead
% -------------------------------------------------------

% =======================================================
% Web Appendix A: Computation of the Critical Constant
% =======================================================
\refstepcounter{webappdummy}\label{Appendix_optimization}

% =======================================================
% Web Appendix B: Proof for the Theoretical Properties
% =======================================================
\refstepcounter{webappdummy}\label{Appendix_Freq_Coverage}

    % B.1: Proof for Theorem 1
    \refstepcounter{websubdummy}\label{Appendix_Th1}

        % B.1.1: Assumptions (no \label in supplement)
        \refstepcounter{websubsubdummy}% placeholder, no label needed

        % B.1.2: Convergence of the point estimator (no \label)
        \refstepcounter{websubsubdummy}% placeholder, no label needed

        % B.1.3: Convergence of the bandwidth (no \label)
        \refstepcounter{websubsubdummy}% placeholder, no label needed

        % B.1.4: Convergence of the critical constant
        \refstepcounter{websubsubdummy}\label{appendix:convergence of lambda}

    % B.2: Proof for Theorem 2
    \refstepcounter{websubdummy}\label{Appendix_Th2}

% =======================================================
% Web Appendix C: Remarks on the Evaluation Criteria
% =======================================================
\refstepcounter{webappdummy}\label{appendix:evaluation}

    % C.1: Posterior simultaneous coverage probability
    \refstepcounter{websubdummy}\label{appendix:PSCP}

    % C.2: Empirical simultaneous coverage rate
    \refstepcounter{websubdummy}\label{appendix:ESCR}

% =======================================================
% Web Appendix D: Supplement to Simulation Studies
% =======================================================
\refstepcounter{webappdummy}\label{Appendix_Simulation}

    % D.1: Hyperparameter specifications
    \refstepcounter{websubdummy}\label{Appendix_Simulation_Hyperparameters}

    % D.2: Sensitivity analysis
    \refstepcounter{websubdummy}\label{Appendix_Simulation_Sensitivity}

    % D.3: Prior and posterior predictive checks
    \refstepcounter{websubdummy}\label{Appendix_Simulation_PPC}

    % D.4: HMC convergence diagnostics
    \refstepcounter{websubdummy}\label{Appendix_Simulation_HMC}

    % D.5: Additional plots
    \refstepcounter{websubdummy}\label{Appendix_simulation_plots}

    % D.6: Additional tables
    \refstepcounter{websubdummy}\label{Appendix_simulation_tables}

% =======================================================
% Web Appendix E: Supplement to Real Data Analysis
% =======================================================
\refstepcounter{webappdummy}\label{Appendix_Real_Data}

% =======================================================
% Web Figures  (sequential numbering)
% =======================================================
\refstepcounter{webfigdummy}\label{fig:NG_prior_posterior_fixAlpha_p3_sigma1}
\refstepcounter{webfigdummy}\label{fig:NG_prior_fixAlpha_p3_sigma1}
\refstepcounter{webfigdummy}\label{fig:NG_prior_posterior_fixedBeta_p3_sigma1}
\refstepcounter{webfigdummy}\label{fig:NG_prior_fixedBeta_p3_sigma1}
\refstepcounter{webfigdummy}\label{fig:NG_J_HMC_p3_sigma1}
\refstepcounter{webfigdummy}\label{fig:acf_p=3_e_sd=1_n=20_iter_1}
\refstepcounter{webfigdummy}\label{fig:density_p=3_e_sd=1_n=20_iter_1}
\refstepcounter{webfigdummy}\label{fig:trace_p=3_e_sd=1_n=20_iter_1}
\refstepcounter{webfigdummy}\label{fig:Prior_predictive_check_p2_n20}
\refstepcounter{webfigdummy}\label{fig:Prior_predictive_check_p3_n20}
\refstepcounter{webfigdummy}\label{fig:Posterior_predictive_check_p2_n20}
\refstepcounter{webfigdummy}\label{fig:Posterior_predictive_check_p3_n20}
\refstepcounter{webfigdummy}\label{fig:3BSCB_Conjugate_p2_sigma1}
\refstepcounter{webfigdummy}\label{fig:3BSCB_Conjugate_p3_sigma1}
\refstepcounter{webfigdummy}\label{fig:3BSCB_HMC_p2_sigma1}
\refstepcounter{webfigdummy}\label{fig:3BSCB_HMC_p3_sigma1}
\refstepcounter{webfigdummy}\label{fig:All_p2_sigma1}
\refstepcounter{webfigdummy}\label{fig:All_p3_sigma1}
\refstepcounter{webfigdummy}\label{fig:theta_boxplot_2_1_20_comparison_4}
\refstepcounter{webfigdummy}\label{fig:theta_boxplot_3_1_20_comparison_4}

% =======================================================
% Web Tables  
% =======================================================
\refstepcounter{webtabdummy}\label{tab:hmc_p3_n20}
\refstepcounter{webtabdummy}\label{tab:ljung_box}
\refstepcounter{webtabdummy}\label{tab:replication}
\refstepcounter{webtabdummy}\label{tab:time}
\refstepcounter{webtabdummy}\label{tab_MPSCP_AR}
\refstepcounter{webtabdummy}\label{tab_ESCR_AR}
\refstepcounter{webtabdummy}\label{table-clinical}
\refstepcounter{webtabdummy}\label{table_real_mean_center}
\refstepcounter{webtabdummy}\label{tab:smallest_dose}